\documentclass{svproc}
\usepackage{url}

\usepackage{bm} 
\usepackage{graphicx}
\begin{document}
\mainmatter              
\title{Excited $\Omega$'s as heavy pentaquarks}

\author{Hyun-Chul Kim\inst{1,2,3}}
%
%
%
\institute{Department of Physics, Inha University, Incheon 22212,
Republic of Korea, \\
\email{hchkim@inha.ac.kr},\\
\and
Advanced Science Research Center, Japan Atomic Energy Agency,
Shirakata, Tokai, Ibaraki, 319-1195, Japan \\ 
\and
School of Physics, Korea Institute for Advanced Study 
  (KIAS), Seoul 02455, Republic of Korea }
\maketitle              

\begin{abstract}
We briefly summarize recent works on the identification of the excited
$\Omega_c$'s found by the LHCb Collaboration. Within the framework of
a pion mean-field approach, the following scenario is the most
favorable: While three of the excited $\Omega_c$'s belong to the
excited baryon sextet, two of them with the smaller decay widths can
be identified as the members of the anti-decapentaplet which is one of
the lowest-lying representations. It implies that these two
$\Omega_c$'s, i.e. $\Omega(3050)$ and $\Omega(3119)$ are most 
probably the exotic heavy pentaquark baryons.  

\end{abstract}
\section{Introduction}
A pentaquark baryon consists of five valence quarks, which was already
mentioned by Gell-Mann~\cite{Gell-Mann} who christened the fundamental  
bulding block of a hadron or the \textit{true} atom a \textit{quark}.     
The LHCb Collaboration reported for the first time the finding of two
heavy pentaquarks that were coined as $P_c(4380)$ and $P_c(4450)$ 
respectively~\cite{Aaij:2015tga,Aaij:2016phn,Aaij:2016ymb,Aaij:2016iza}.
The valence quark content of these heavy pentaquarks is given
$uudc\bar{c}$. They can be considered as resonances involving the
$J/\psi$ $(c\bar{c})$  and the proton ($uud$). Yet another interesting 
finding of five excited $\Omega_c^*$'s was announced also by the LHCb 
Collaboration~\cite{Aaij:2017nav}, four of which were confirmed by
the Belle Collaboration~\cite{Yelton:2017qxg}. 

One can naturally classify them as the members of the excited baryon
sextets, since there exist the five sextet representations, once one
of the valence quarks are excited to the level with the orbital
angular momentum $l=1$. In the present talk, we will show that this
classification is inconsistent with the experimental data and will
propose that two of them can be regarded as members of the
baryon anti-decapentaplet
($\overline{\bm{15}}$)~\cite{Kim:2017jpx,Kim:2017khv}, 
which appear as one of the lowest representations for singly heavy
baryons. By the member of the baryon anti-decapentaplet we mean that
it is a heavy pentaquark baryon.  

The general theoretical framework we employ in this work is a pion
mean-field approach, which can be also called as the chiral
quark-soliton model ($\chi$QSM). Mean-field approximations have
enjoyed simple but clear understanding of various physical problems in
different branches of physics and have been used even in different
fields such as applied mathematics,computer sciences, etc. The main
idea of the pion mean-field approach is that the quantum fluctuation
of the pion field, which can be regarded as the $1/N_c$ corrections in
the large $N_c$ expansion, can be
suppressed~\cite{Witten:1979kh}. Then, the pion mean 
field arises from the classical solution of the equation of motion for
the pion around the saddle
point~\cite{Diakonov:1987ty}. This $\chi$QSM was successfully used to
describe numerous properties of the nucleon and SU(3) hyperons,
including their static properties and form
factors~\cite{Christov:1995vm} and parton distributions. Recently, the
$\chi$QSM was extended to the description of singly heavy
baryons~\cite{Yang:2016qdz,Kim:2018xlc,Kim:2018nqf} 
(see also a recent review~\cite{Kim:2018cxv}),
being motivated by Ref.~\cite{Diakonov:2010tf}.  

In the present talk, we want briefly explain how the newly-found
excited $\Omega_c^0$'s~\cite{Aaij:2017nav} can be classified uniquely
within the framework of the pion mean-field approaches: Three of the
excited $\Omega_c$'s can be naturally understood as the members of the
excited baryon sextet whereas two of them, which have relatively
smaller decay widths, should belong to the ground baryon
anti-decapentaplet. If this scenario turns out true, then the charged
$\Omega_c^*$'s in the invariant-mass $\Xi_c^+ K^0$ and $\Xi_c^0 
K^-$ channels will be observed.  

\section{Singly heavy baryons as a system of
 $N_c-1$ valence quarks in pion mean fields}
\begin{figure}[htp]
\centering
\includegraphics[scale=0.18]{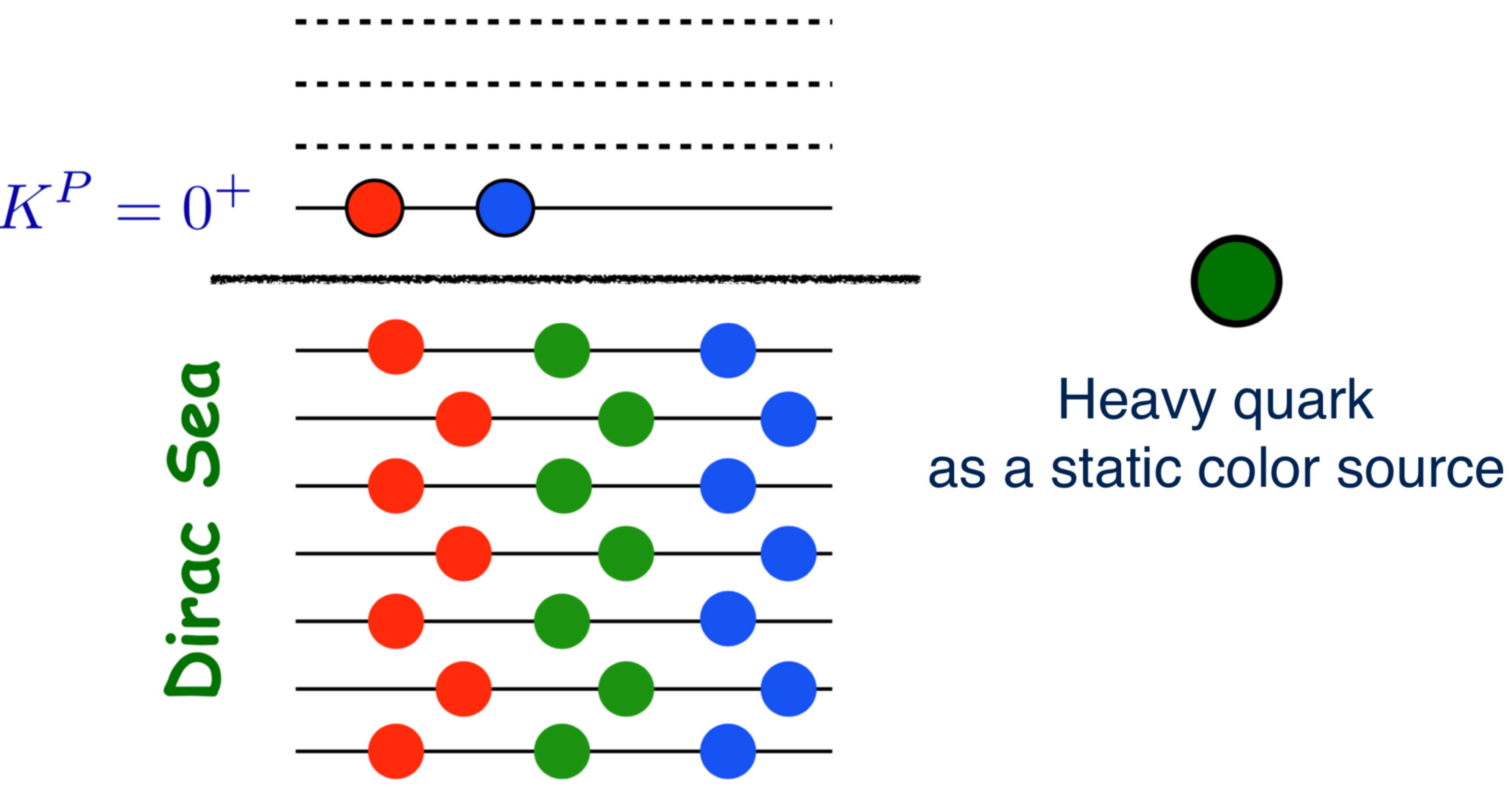}
\caption{Lowest-lying singly heavy baryons.} 
\label{fig:1}
\end{figure}

\begin{figure}[htp]
\centering
\includegraphics[scale=0.22]{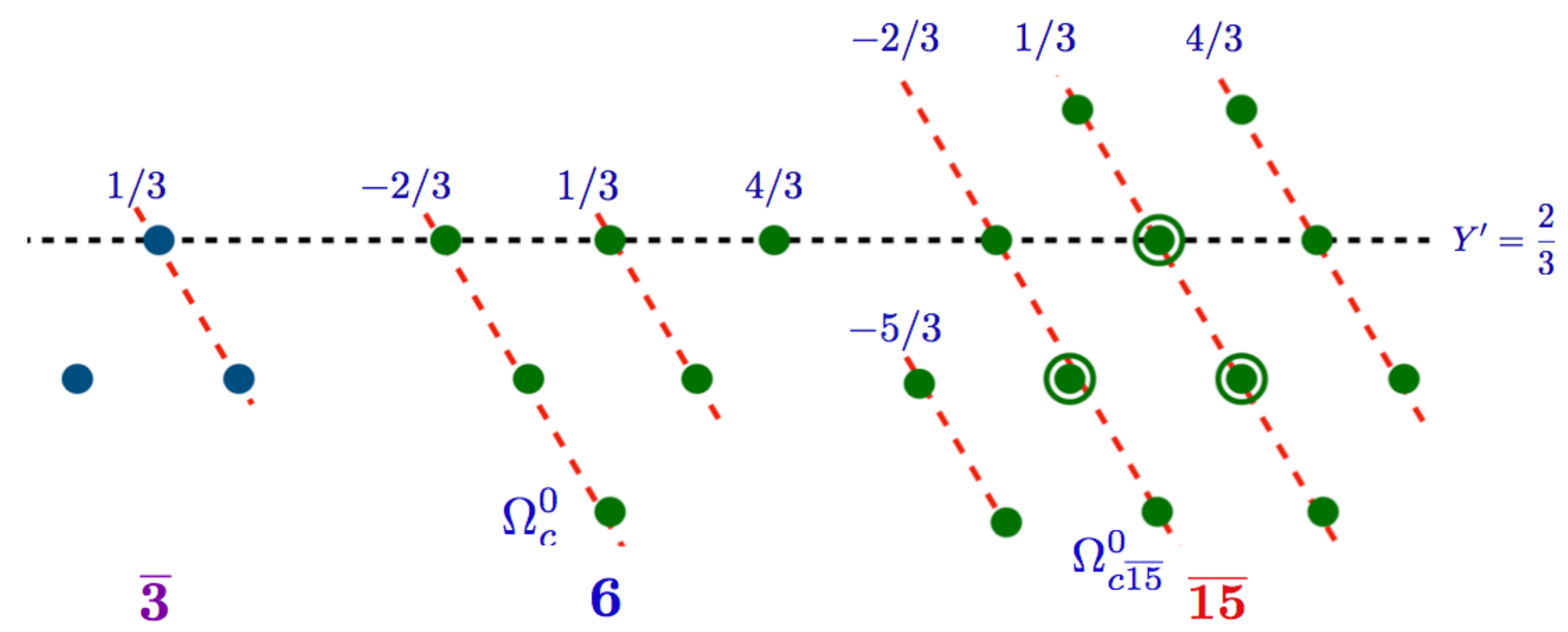}
\caption{Allowed representations of the singly heavy baryons} 
\label{fig:2}
\end{figure}

In the pion mean-field approach, a light baryon in a low-lying
representation can be regarded as $N_c$ valence quarks
self-consistently bound by the pion mean fields that are produced by
the valence quark themselves. The process is nothing but a 
well-known Hartree approximation from atomic and nuclear physics. 
On the other hand, a singly heavy baryon consists of the $N_c-1$
valence quarks while the heavy quark is regarded as a spectator,
i.e. a mere static color source in the limit of the infinitley heavy
quark mass ($m_Q\to\infty$), as  depicted in Fig.~\ref{fig:1}.  
Thus, the singly heavy-quark baryons with $N_c-1$ quarks have the
constraint on the quantization, i.e. the right hypercharge $Y'=(N_c-1)/3$ 
instead of $Y'=N_c/3$. In this case, allowed representations must
include states with the value of $Y'$ when 
$N_c=3$: the antitriplet ($\overline{\bm{3}}$), the sextet ($\bm{6}$),
and the anti-decapentaplet ($\overline{\bm{15}}$) as illustrated in
Fig.~\ref{fig:2}.  
The isospin $\bm{T}$ of the ground states with $Y'=(N_c-1)/3$ is also
constrained by the relation $\bm{T}+\bm{J}=\bm{K}=\bm{0}$, where
$\bm{J}$ denotes the soliton spin that must be coupled to $\bm{T}$ to 
give the grand spin $\bm{K}$.  The ground-state heavy baryons must
have $K=0$ because all the valence quarks lie in the state $K^P=0^+$
with parity $P$. By these selection rules, the $\overline{\bm{3}}$ has
the soliton spin $0$ whereas the $\bm{6}$ has spin $1$.  The
$\overline{\bm{15}}$ has both spin 0 and 1. The soliton spin being
coupled to the heavy-quark spin $1/2$, the singly heavy baryons can be
finally constructed as the antitriplet with spin $1/2^+$, the sextet with
spin $1/2^+$ and $3/2^+$ that will be split by a hyperfine
interaction, and the anti-decapentaplet with $1/2^+$ and $3/2^+$.

\begin{figure}[htp]
\centering
\includegraphics[scale=0.18]{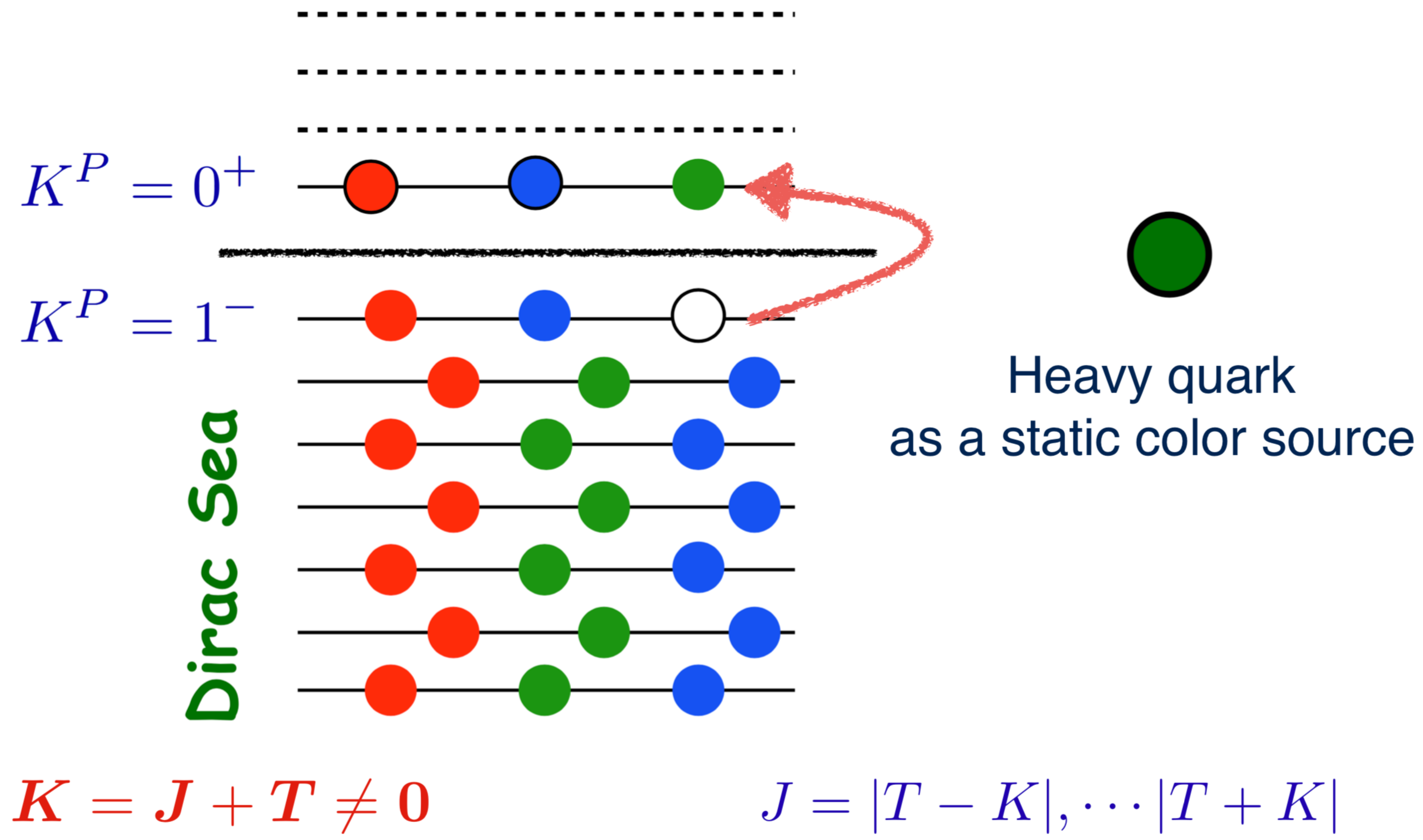}
\caption{Excited heavy baryons}
\label{fig:3}
\end{figure}

While a nucleon or a hyperon is excited by triggering a lowest-lying
valence quark to hop into the next excited level with $K^P=1^-$, 
the most favorable way of describing an excited singly heavy baryon is
to excite a sea quark in the $K^P=1^-$ level to an unoccupied
$K^P=0^+$ level to fill up, as demonstrated in
Fig.~\ref{fig:3}.
Using the quantization rule $\bm{K}=\bm{J}+\bm{T}$, we can classify
the allowed representations for the excited baryons. Since
$K=1$, we first consider the simplest case $T=0$ that gives $J=1$,
which belongs to the excited baryon anti-triplet. Combining a charm
quark with spin 1/2, we find there are \textit{two} excited
antitriplet representations respectively with spin 1/2 and 3/2, which
will be split by a certain hyperfine interaction. When $T=1$, the
allowed $J$ of the soliton can be $J=0,\,1,\,2$. Being coupled to the
charm quark, \textit{five} different sextet representations appear.   
\section{Two scenarios for identifying excited 
$\Omega_c$'s}
Very recently, Ref.~\cite{Kim:2017jpx} have scrutinized
the spectrum of the excited $\Omega_c$'s reported by the LHCb
Collaboration within the framework of the $\chi$QSM.  Since five
excited $\Omega_c$'s were found, it is very natural to regard them as
the members of the excited baryon sextets, since we have exactly the
five sextet representations as shown previously. Thus, we first
examined how these five sextet representations are split. 

Expressing the mass splitting between $J=0$ and $J=1$ states as 
$\Delta_1$ and that between $J=1$ and $J=2$ states as $\Delta_2$, we
find a relation $\Delta_2=2\Delta_1$ within the present
approach~\cite{Kim:2017jpx}. This relation is the robust one that will
play a role of a touchstone to judge which scenario is more plausible
and viable. In addition,  one needs to find the hyperfine splittings 
between the sextets. Assuming that $\Lambda_c^+(2595)$ and
$\Xi_c(2790)$ belong to the excited antitriplet with $J^P=1/2^-$
whereas $\Lambda_c(2625)$ and $\Xi_c(2818)$ are the members of that
with $3/2^-$, we can fix the parameter $\kappa'/m_c\approx
30\,\mathrm{MeV}$ for the hyperfine splitting. The hyperfine splitting
between the sextet with $1/2^-$ and $3/2^-$ is given by the same
$\kappa'/m_c$ as in the antitriplet, but that between the sextet with
$3/2^-$ and $5/2^-$ is written by $5\kappa'/3m_c$. 

In the first scenario, we assume that the five excited
$\Omega_c$'s found by the LHCb belong to each baryon sextet. As
mentioned previously, it seems a natural scenario, since there are
five representations of the sextet. However, this first scenario leads
to the discrepancies that $\kappa'/m_c$ should be at least two times
smaller than that determined from the baryon antitriplet and the relation
$\Delta_2=\Delta_1$ is badly broken within this assumption. In
addition, the sum rules between the $\Omega_c^*$ masses are also
unfavorably violated~(see Ref.~\cite{Kim:2017jpx} for details).  

Thus, we propose the second scenario in which we assert that three
of $\Omega_c^*$'s together with certain bump structures above 3.2 GeV
in the LHCb data belong to the five baryon sextet while two of them
($\Omega_c(3050)$ and $\Omega_c(3119)$),
which have smaller decay widths, are associated with the
anti-decapentaplet. First of all, the relation $\Delta_2=2\Delta_1$ is
almost perfectly satisfied and the value of the hyperfine splitting is
obtained to be $\kappa'/m_c\approx 25$ MeV which is much closer to
that determined from the baryon anti-triplet in comparison with the
first scenario. Since the $\overline{\bm{15}}$ is one of the lowest
representations for the singly heavy baryons, we have
$\bm{J}=\bm{T}=\bm{1}$. Being coupled to the heavy-quark spin, there
exist two different $\overline{\bm{15}}$ representations with $1/2^+$
and $3/2^+$, respectively. We assign spin $1/2$ to $\Omega_c(3050)$
whereas $\Omega_c(3119)$ is allocated to a spin $3/2$
state. Surprisingly, the hyperfine mass 
splitting between $\Omega_c(3050)$ and $\Omega_c(3119)$ is $69$ MeV,
which is in excellent agreement with the value obtained from the
ground-state baryon sextet: $\kappa/m_c=(68.1\pm 1.1)$
MeV~\cite{Yang:2016qdz}. The results indicate that the second scenario
successfully classifies the five excited $\Omega_c$'s observed by the
LHCb Collaboration. 

Since the $\Omega_c(3050)$ and $\Omega(3119)$ have a rather smaller
decay widths than the other $\Omega_c$'s, we have to show that the
$\chi$QSM model can explain it. In Ref.~\cite{Kim:2017khv}, the strong
decay widths of the gound-state singly heavy baryons and baryon
anti-decapentaplet. Since all the dynamical parameters for the
axial-vector transitions were already fixed by using the semileptonic
decay constants of the SU(3) hyperons~\cite{Yang:2015era}, one can
immediately compute the strong decay widths of the anti-decapentaplet
$\Omega_c$'s. In flavor SU(3) symmetry, we were able to reproduce very
well the existing experimental data on the strong decay widths of the
transition from the ground-state baryon sextet ($\bm{6}_{1/2}$ and
$\bm{6}_{3/2}$) to the ground-state antitriplet ($\overline{\bm{3}}$)
~\cite{Kim:2017khv}. Having shown that the present approach
successfully describe the known experimental data on the heavy-baryon
strong decays, we continued to compute the strong decays of
$\Omega_c(3050)$ and $\Omega(3119)$. There are three decay modes for
$\Omega_c(3050)$, i.e. $\Omega_c(\overline{\bm{15}}_1,1/2)\to
\Xi_c(\overline{\bm{3}}_0,1/2)+ K$, $\Omega_c(\overline{\bm{15}}_1,1/2)\to
\Omega_c(\bm{6}_1,1/2)+ \pi$, and
$\Omega_c(\overline{\bm{15}}_1,1/2)\to \Omega_c(\bm{6}_1,3/2)+ \pi$,
whereas $\Omega_c(3119)$ can decay into four different modes:
$\Omega_c(3050)$, i.e. $\Omega_c(\overline{\bm{15}}_1,3/2)\to 
\Xi_c(\overline{\bm{3}}_0,1/2)+ K$,  $\Omega_c(\overline{\bm{15}}_1,3/2)\to 
\Xi_c(\bm{6}_1,1/2)+ K$, $\Omega_c(\overline{\bm{15}}_1,3/2)\to
\Omega_c(\bm{6}_1,1/2)+ \pi$, and $\Omega_c(\overline{\bm{15}}_1,3/2) 
\to \Omega_c(\bm{6}_1,3/2)+ \pi$. We obtained the total decay widths
of $\Omega_c(3050)$ and $\Omega_c(3119)$ as 
$\Gamma_{\Omega_c(3050)}=0.48$ MeV and $\Gamma_{\Omega_c(3119)} =
1.12$ MeV, respectively. Compared to the experimental data
$\Gamma_{\Omega_c(3050)}=(0.8\pm0.2\pm0.1)$ MeV and
$\Gamma_{\Omega_c(3119)} =(1.1\pm0.8\pm0.4)$ MeV, the present results
are in very good agreement with the data. Moreover, the smallness of
the $\Omega_c(3050)$ and $\Omega_c(3119)$ decay widths is clearly 
explained by large cancellation between the leading order and the 
next-to-leading order corrections in the large $N_c$ expansion. In
fact, the widths of the pentaquarks vanish in the nonrelativistic
limit of the present model, which is the very same as the case of the
light pentaquark $\Theta^+$. The smallness of the decay widths
is one of the typical characteristics of a pentaquark states. Results
for other members of the $\overline{\bm{15}}$, we refer to 
Refs.~\cite{Kim:2017jpx,Kim:2017khv}.        
\section{Conclusion and outlook}
As noted previously, if one classifies the $\Omega_c(3050)$ and
$\Omega_c(3119)$ as the members of the baryon anti-decapentaplet,
their isospins should be equal to one. It implies that there must be
three $\Omega_c$ in each $\overline{\bm{15}}$ representation with
different charges. Thus, the $\Omega_c^+$ and $\Omega_c^-$ should
exist. In fact, both the $\Omega_c(3050)$ and $\Omega_c(3119)$ were
observed in the invariant mass of the $\Xi_c^+K^-$ channel. Thus,
there is a chance for other isospin members to be found in the charged
channels, i.e. $\Xi_c^+K^0$ or $\Xi_c^0 K^-$ decay
channels. Corresponding experiments can be performed by the LHCb and
Belle II Collaborations. We anticipate possible findings of the
charged $\Omega_c$'s in near future. The strong decays of the heavy
baryons and excited $\Omega_c$ are under investigation with flavor
SU(3) symmetry breaking taken into account. 
\section*{Acknowledgments} 
The author is very grateful to M. V. Polyakov, M. Prasza{\l}owicz, and 
Gh.-S. Yang for fruitful collaboration over decades. The present talk
was based on a series of works carried out together with them. 
He is also grateful to P. Gubler, T. Maruyama and M. Oka for useful 
discussions.  He wants to express his gratitude to the members of the
Advanced Science Research Center (ASRC) at Japan Atomic Energy Agency
(JAEA) for the hospitality, where part of the present work was done. 
This work was supported by the National Research Foundation of Korea
(NRF) grant funded by the Korea government (No. NRF-2018R1A2B2001752).   

%

\end{document}